\documentclass{aa}
\usepackage{graphicx}
\usepackage{txfonts}
\def\lya{Ly$\alpha$ }

\begin{document}
   \title{The physical properties of Ly$\alpha$ emitting galaxies: not just 
primeval galaxies? }

    
   \author{   L.Pentericci\inst{1}
             \and
          A. Grazian \inst{1}
          \and 
          A.Fontana  \inst{1}
          \and 
M. Castellano  \inst{2}
  \and 
E. Giallongo       \inst{1}
  \and 
S. Salimbeni   \inst{1,3}
 \and 
P. Santini \inst{1}
    }

   \offprints{L.Pentericci}

   \institute{INAF - Osservatorio Astronomico di Roma, Via Frascati 33,
I--00040, Monte Porzio Catone, Italy
 \and Dipartimento di Fisica, Universita$'$ di Roma ``La Sapienza'', P.le
A. Moro 2,00185, Roma, Italy 
\and 
University of Massachusetts, Department of Astronomy, 710
North Pleasant Street, Amherst, MA 01003, USA
           }
   \date{}


  \abstract
   {}
{We have analyzed a sample of 
Lyman Break Galaxies from z$\sim 3.5 $ to z$\sim 6$ 
selected from the  GOODS-S field  as B,V and i-dropouts, and with spectroscopic
 observations showing that they have the \lya\ line in emission.
Our main aim is  to  investigate  their physical properties 
and their dependence on the emission line characteristics, to shed light on the relation between galaxies 
with Ly$\alpha$ emission and the general LBG population.
}
   { The objects were selected from their optical continuum colors and then 
spectroscopically  confirmed by the GOODS collaboration and other  campaigns.
From the public spectra we derived the main properties 
of the \lya\ emission such as total
 flux and rest-frame  EW.
We then  used complete photometry, 
from U band to mid-infrared from the GOODS-MUSIC 
database, and through standard spectro-photometric techniques we derived 
 the physical properties of the galaxies, such as total stellar mass, stellar ages, 
star formation rates and  dust content.
Finally we investigated the relation between emission line and physical properties.
}
   {Although most galaxies are fit by  young stellar populations, 
a small but non negligible fraction
has SEDs that cannot be well represented by young models and 
require considerably older stellar component, up to $\sim 1$Gyr. 
There is no apparent relation between age and EW: some of the oldest galaxies have large
line EW, and  should be also selected in narrow band surveys.
Therefore not all Ly$\alpha$ emitting galaxies
are primeval galaxies in the very early stages 
of formation, as is commonly assumed.
\\
We also find a large range of stellar populations, with masses from $5\times 10^8 M_\odot$ to
 $5 \times 10^{10} M_\odot$  and SFR from few to  60$M_\odot yr^{-1}$.
Although there is no net correlation between mass and EW,
we find a significant  lack of massive galaxies with large EW, which could be explained if the most massive galaxies
were either more dusty and/or contained more neutral gas than less massive objects.
\\
Finally we find that more than half of the galaxies contain small but non negligible amounts of dust:
the mean  E(B-V) derived from the SED fit and the  EW are well
correlated, although with a large scatter, as already found at lower redshift.
}
   {}
\maketitle

\section{Introduction}


In the last few years large samples of high redshift star forming galaxies have been found 
at increasingly larger distances (e.g. Bouwens et al. 2006, Iye et al. 2006).
Their photometric
and physical properties as well as  spatial distribution have been extensively studied.
The majority of these galaxies are detected on the bases of their typical broad band colors, 
given by the characteristics breaks (Lyman break, Lyman limit) 
that fall at different redshift into the different bands (the Lyman break galaxies LBGs 
e.g. Steidel et al. 1996). Alternatively many galaxies have been found by means
of their bright \lya\ emission, in particular at redshift $> 3$:  Ly$\alpha$ emitters (LAEs) 
are selected through ultra-deep narrow band (NB) and by contrast 
to a nearby broad band image, as initially proposed  by  Cowie \& Hu (1998) and then used by many others
(e.g. Iye et al. 2006, Ouchi et al. 2005, Fujita et al. 2003).
This technique  tends to select galaxies with relatively 
 faint continuum emission and large line equivalent width (EW).
\\
Each of the two methods suffers from a different selection bias: the two resulting populations of galaxies  
overlap partially and the  relationship  between them is not clear.
Most authors have shown  that LAEs are, on average, smaller and younger galaxies as compared to the LBGs 
population (e.g.\ Finkelstein et al. 2007, Gawiser et al. 2007, Pentericci et al. 2007 hereafter P07 ).
 Because the \lya\ line is easily suppressed by dust, \lya\ emitters are often characterized as extremely 
young galaxies, experiencing their initial phase of  star formation in essentially dust free environments (e.g. 
Gawiser et al. 2007).
However the different behavior of \lya\ and continuum photons in interacting with dust, makes it possible also for older galaxies to exhibit \lya\ in emission, as predicted e.g. in the models of Haiman \& Spaans (1999). Therefore LAEs (or a fraction of them) could also represent an older population with active star forming regions, where the gas kinematics can favor the escape of \lya\ emission.
This scenario is partially  supported  by the results of Shapley et al. (2003). 
In addition  Lai et al. (2007) found that some high redshift \lya\ galaxies  could be consistent with hosting relatively old stellar population.
\\
The fraction of Lyman break galaxies that are also LAEs is
 also still under debate.
Some authors have recently claimed a deficiency  of 
 bright galaxies with large  EW in deep samples of Lyman break galaxies, indicating that the fraction of \lya\ emitters amongst LBGs might change abruptly 
with UV luminosity  
(Ando et al. 2006).
On the other hand  Shimasaku et al. (2006) argue that at redshift 6 the 
fraction of   LBGs with    $EW> 100 \AA$ is 80\%.
They claim that the fraction of \lya\ emitters 
amongst LBGs is a strong function of redshift.
This is at variance with  the results of Dow-Hygelund et al. (2007, see also Stanway et al. 2007), who find that the fraction of \lya\ emitters in galaxies at $z=6$ is $\sim 30\%$, in total accordance 
with what found by Shapley el al. (2003) for z=3 Lyman break galaxies, indicating
that there is no strong evolution  between $z\sim 3$  and $z\sim 6$. 
\\
Clearly, it  is worthwhile understanding  these trends and 
the real relation between galaxies with Ly$\alpha$
emission and the general LBG population,
 so that properties of the overall high-redshift galaxy population, such as the total stellar
mass density, can be better constrained.
\\
An important limitation of the LAE technique 
is that it tends  to select  galaxies with extremely faint continuum:
 therefore most of the LAEs are not
detected, or just barely detected,in the broad bands.
Thus the analysis of all physical properties that are usually derived
from a modeling of the multi-band spectral energy distributions, 
such as stellar masses and ages is extremely difficult.
In most cases the determination of these properties for the individual objects is not possible and one has to rely on the analysis of stacked data
 (e.g.  Lai et al. 2006).
In other cases, the analysis is limited to very small sub-sample of LAEs that are detected in the continuum  (e.g. Finkelstein et al. 2007).
For example, it has been shown that to reliably estimate the masses at redshift larger than $z\sim 2.5$,
the inclusion of near and mid IR bands is essential (e.g. Fontana et al. 2006,  hereafter F06) to
constrain the values and reduce the uncertainties.
In this context, Lai et al. (2007) attempted to constrain the
stellar population of $z=3.1$ LAEs selected from the Extended CDFS, using Spitzer data.
However of their initial  sample of 162 LAEs, only 18 galaxies were  
detected in the IRAC 3.6$\mu$ channel, 
and therefore had reliable individual  mass estimates.
For the rest of the sample, a limit in  mass was 
determined from a  stacking analysis, obtaining  only information on the average properties. This makes it hard  to analyze 
the correlation and trends between the various properties.
\\
For the reasons detailed above and to ensure that we can derive the physical
properties of \lya\ emitting galaxies, 
we chose to  analyze  galaxies exhibiting Ly$\alpha$ in emission starting from a sample of LBGs.
Thanks to the large area coverage of GOODS, in this way we can assemble a sample with a large enough number of galaxies, 
and a wide range of \lya\ properties, from bright emission lines to 
absorption. 
 We then selected only 
those exhibiting \lya\  in emission.
 We do not include in this analysis those LBGs showing  
the \lya\ line in absorption,
since our main aim is to compare the physical properties of our galaxies
 to those of NB selected samples. 
Furthermore the fraction of LBGs with Ly$\alpha$ in absorption decreases with redshift, since it becomes progressively harder to identify 
\lya\ absorbers. Therefore to 
keep the sample complete for \lya absorbers 
we should have stopped at a lower redshift (as in P07) and 
the sample would have been smaller. 
\\
The paper is organized as follows: in Section 2 we outline the sample selection and  in Section 3 
we describe how the physical properties were derived from the multiwavelength observation  
and the properties of the \lya\ emission from the spectroscopic observations.
In Section 4 we analyze the various trend of physical properties with the characteristics of the line emission, 
and finally in Section 5 we discuss our results in the context of various proposed scenarios.
  \\
All magnitudes are in the AB system (except where otherwise stated)
and we adopt the
$\Lambda$-CDM concordance cosmological model ($H_0=70$ , $\Omega_M=0.3$ and
$\Omega_{\Lambda}=0.7$).

\section{ Sample selection}
We used a revised version of the  GOODS-MUSIC z-selected sample of galaxies that will be presented in Santini et al. (2008 submitted). 
The 
limiting magnitude $z$ that varies for different areas
of the field (see Grazian et al. 2006a for more details) but is typically around $z\sim 26$ (AB scale).
The  color-color selection criteria have been  presented by  Giavalisco et al. (2004) and were used for the ESO spectroscopic campaign of the GOODS south field. We adopt the  B, 
 V and i dropout selections, which give samples of galaxies at redshift approximately 
between 3.5 and 4.5 (B dropouts) between 4.5 and 5.5 (V dropout) and above 5.5 (i dropouts).
All known AGNs have been removed by identifying the objects that have either 
broad line emission or X-ray detections (see Santini et al. 2008 for more details).
This way we assembled a sample of several hundreds LBGs:
we then selected  those that were observed spectroscopically.
Most of the spectra were taken within the GOODS-FORS2 spectroscopic campaign (Vanzella et al. 2005, 2006, 2008)  and the 1D spectra were retrieved directly 
from the public GOODS database. 
The GOODS team has classified galaxies according to the presence of
Ly$\alpha$ in emission or absorption.
We further checked their classification and we retained only those 
galaxies with Ly$\alpha$ in emission (few spectra present the line both in emission and absorption and were also included in the final sample).
We included  galaxies with spectroscopic classification A and B  (respectively
secure and probable redshift determination)
we  also retained  
those galaxies with spectroscopic classification  C (tentative redshift determination) if
our independently determined  photometric redshift  
(from 14 bands photometry, see G06 
 for more details) is in agreement with the spectroscopic one.
\\
\begin{figure}
\includegraphics[width=9cm]{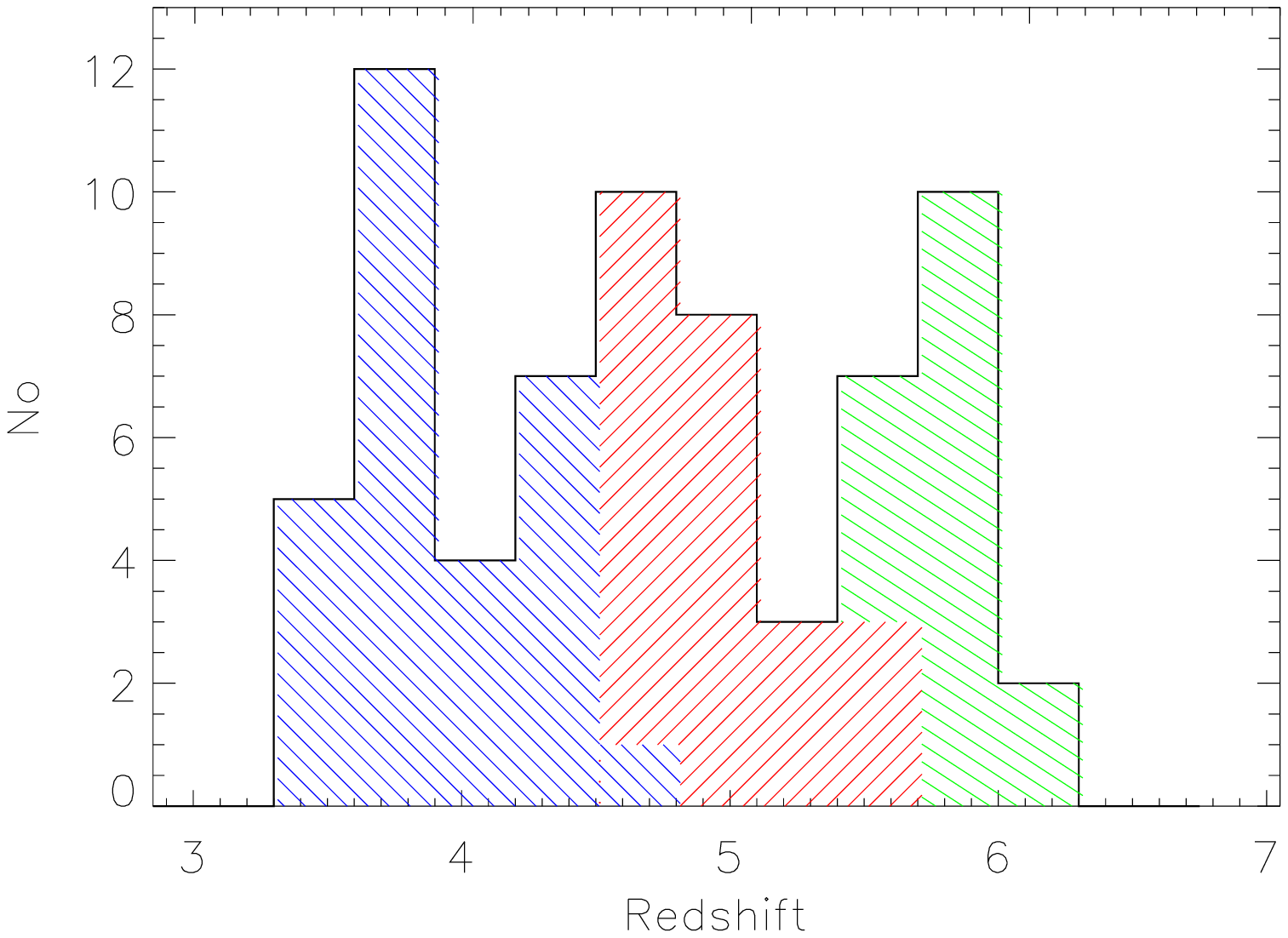}
\caption{The redshift distribution of LBG galaxies in our sample: blue are the B dropouts, red are the V dropouts and 
green are the i-dropouts (see text for details).}
\includegraphics[width=9cm]{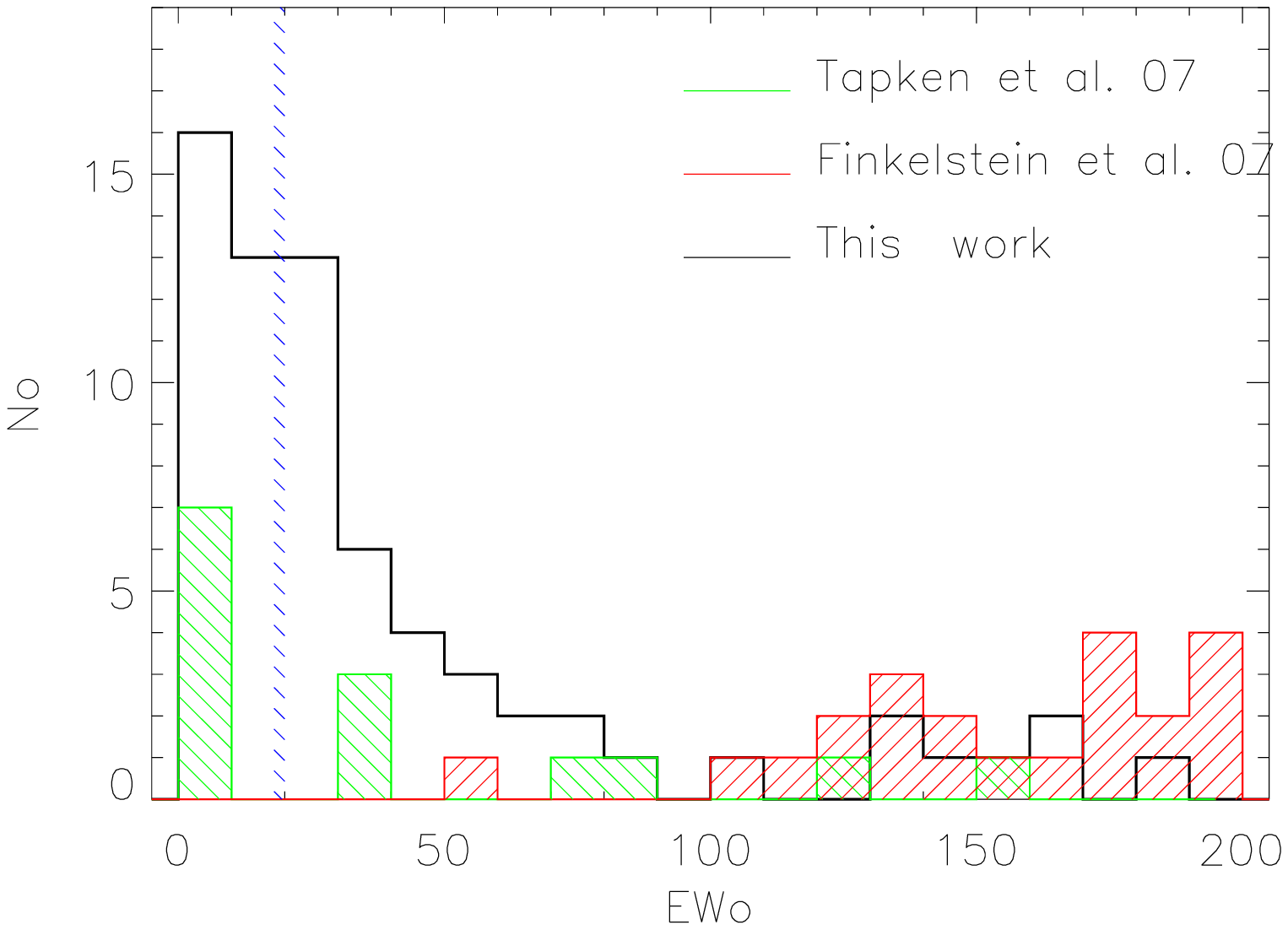}
\caption{The distribution of  \lya\ EW (rest-frame) as determined from the 1d spectra (values are not  corrected for IGM absorption). The black histogram is the sample of LBGs presented in this work, the red dashed area is the sample of LAEs by Finkelstein et al. (2007) and the green dashed area represents the sample of LBGs by Tapken et al.(2007). The vertical blue line at 20 \AA\ indicates the limit for NB selected LAEs.
}
\label{fig1}
\end{figure}
Few further spectra
were retrieved from the GOODS-VIMOS database (Popesso et al. 2008).
For these spectra we followed the same procedure as above,
but we checked all of them  for consistency with our photometric redshift, regardless of the spectral quality.
\\
Finally few galaxies  had published HST/ACS grism spectroscopy either  from the GRAPES observing program 
(Malhotra et al.) or  from preliminary PEARS results (Pirzkal et al. 2007). 
For these galaxies the spectra in electronic format are not public,
but we retrieved all relevant information on the emission lines directly 
from the papers.
\\
The final sample consists of 68 spectroscopically confirmed LBGs: in Figure 1
we present the redshift distribution for B, V and i dropouts.
\\
\section{Physical and spectral properties}
The main physical properties of the galaxies such as total stellar mass, 
continuum-based star formation rate, stellar age,
dust extinction  E(B-V) and so on, were obtained trough 
a spectral fitting technique which has been developed in previous papers
(Fontana et al. 2003, F06), and is similar
to those adopted
by other groups in the literature (e.g. Dickinson et al. 2003, Drory et al. 2004).
Briefly, it is based on a comparison between the observed multicolor
distribution of each object and a set of templates, computed with standard
spectral synthesis models. We used both the 
Bruzual \& Charlot (2003) 
models for a consistent comparison with previous work, and the 
new  Charlot \& Bruzual (2007) models that include more 
recent calculations of evolutionary tracks of TP-AGB stars of different mass and metallicity.

The models were chosen to broadly encompass the variety of star--formation histories,
metallicity and extinction of real galaxies. For purposes of comparison with
previous research, we used the Salpeter IMF, ranging over a set of
metallicities (from $Z=0.02 Z_\odot$ to $Z=2.5 Z_\odot$) and dust
extinction ($0<E(B-V)<1.1$, with a Calzetti or a Small Magellanic Cloud
extinction curve). Details
are given in Table 1 of Fontana et al.(2004).
For each model of this grid, we
computed the expected magnitudes in our filter set, and found the
best--fitting template with a standard $\chi^2$ minimization. The
stellar mass and other best--fit parameters of the galaxy,
such as SFR estimated from the UV luminosity  and corrected for dust obscuration (with a typical correction factor
of $A_V \sim 0.4$), age, $\tau$ (the star formation e-folding timescale),
metallicity and dust extinction, are fitted simultaneously
to the actual SED of the observed galaxy.
The derivation of these parameters is explained in detail
in the above paper and in F06, where
the uncertainties are also discussed.
In particular the
stellar mass generally turns out to be the least sensitive to
variations in input model assumptions; the extension of the
SEDs to the IRAC mid-IR data
tends to reduce considerably the formal uncertainties
on the derived stellar masses.
On the other hand, the physical parameter with highest associated uncertainty is the metallicity, 
given that the models are strongly degenerate when fitting broad-band SEDs.

To characterize the \lya\ emission we  determined the following properties:  the 
line equivalent width (EW), the width at half maximum (FWHM) and the 
total line flux. Where possible we used the flux and 
wavelength calibrated spectra provided by the 
GOODS team. For details on the reduction and 
calibration process and the involved uncertainties of 
the FORS2 spectra we refer to Vanzella et al. (2005, 2006, 2008).
\\
The total line flux and the equivalent width were measured from the spectra 
using as reference continuum a measure from the region immediately red-ward of the line.
In some case when  this was particularly noisy the uncertainties 
(especially on the EW) are quite high. When no continuum is observed 
in the spectrum, the EW are lower limits.
 The measured EW were then divided by (1+z) to determine 
the rest-frame values.
In Figure 2 we show the histogram of the EWs for all galaxies: the distribution
 is peaked at small values and  spans  the 
range from 0 to about 100 \AA, with few objects having EW above 100\AA. 
In the plot, we  also show the distribution of other galaxy samples already discussed in the introduction:
the red histogram represents  the 22 LAEs from Finkelstein et al. (2007) 
 for which masses were determined in a reliable way.
The green  histogram represents the sample of 14 LBGs by Tapken et al. (2007), selected with similar criteria 
from the FORS  deep field, 
which  spans a similar range of redshifts and has an  EW distribution comparable to our sample.
\\
In the plot we also indicated the 20\AA\  rest-frame EW  that is used by most authors to select LAEs 
from deep narrow band surveys. In our sample, 38 of 68
galaxies have EW larger than this value.
This is consistent with the  statistics of LBGs at z$\sim 3$ 
of Shapley et al. (2003), who show that 
of all LBGs, 50\% have  \lya\ in emission  and half of those 
(i.e. $\sim$25\% of the total) have rest-frame EW exceeding 20\AA.
 In the rest of the paper we will refer to the 38 galaxies with $EW > 20\AA\ $
as the NB subsample.
\\
To determine the intrisic emission line flux we  corrected the values measured 
from the spectra 
for IGM absorption.
Ly$\alpha$ sits right at a step function in the Madau (1995) IGM treatment, 
so the amount of attenuation applied to Ly$\alpha$ depends strongly on the exact wavelength position of the Ly$\alpha$ line. The accepted interpretation is that the internal kinematics of a given galaxy will result in half of the 
\lya\  flux coming out slightly blue of the rest wavelength, and half slightly red. 
This results in the characteristic asymmetric profile of the Ly$\alpha$ observed in many spectra, where the blue side is truncated.
 We therefore applied the Madau (1995) 
prescription assuming that only a half of the flux is attenuated,  and  we derived the intrinsic flux. As pointed by several authors (e.g. Santos 2004, Dijkstra et al. 2007) this is a simplistic approach. In particular the IGM around a galaxy is probably overdense and has peculiar velocities: these models predict that the fraction of \lya\ flux 
 that is trasmitted might be lower 
than 0.5, and/or might fluctuate between galaxies. 
\\
Finally we measured the line FWHM  from the spectra by fitting a Gaussian
 to the red part of the spectra (that should be unabsorbed).
Given that the spectra were not taken with  
high resolution set-up, this is probably a simplistic approach.
The measured FWHM were then deconvolved by the resolution 
of the spectroscopic set-up ($R=660$).
\\
For the small sample of objects that were observed by GRAPES,  the EWs 
were determined from the values of narrow band  \lya\  
and continuum flux given in the relevant papers (Pirzkal et al. 2007).
In these cases no value for the FWHM are available.
No significant correlations were found between FWHM and other physical  quantities, probably because  the resolution is not adequate for line shape analysis (see also Tapken et al. 2004).
Therefore we will not discuss  them futher in the paper.

\section{Results:}
\subsection{Old Ly$\alpha$ emitters? }
\begin{figure}
\includegraphics[width=9cm]{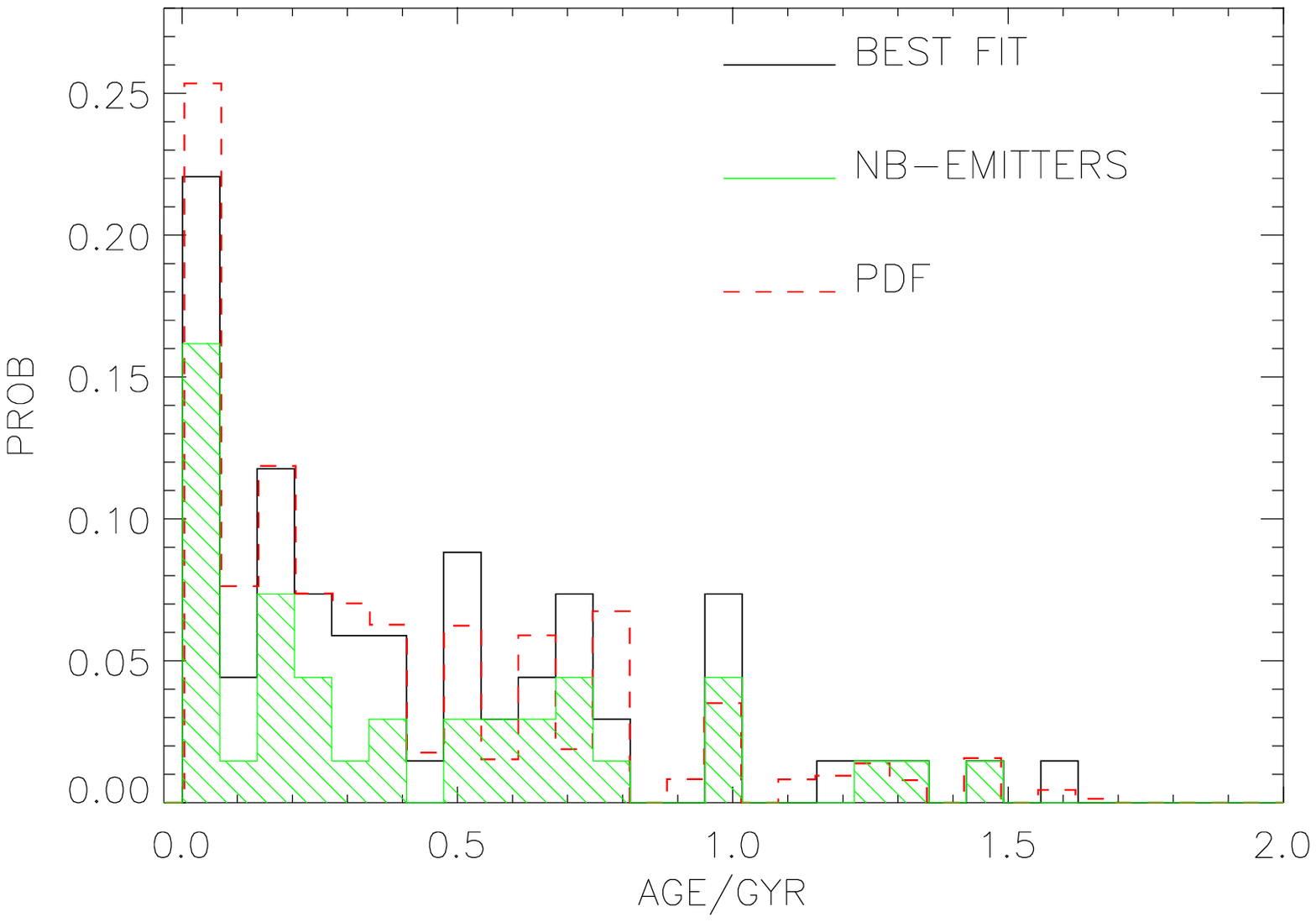}
\includegraphics[width=9cm]{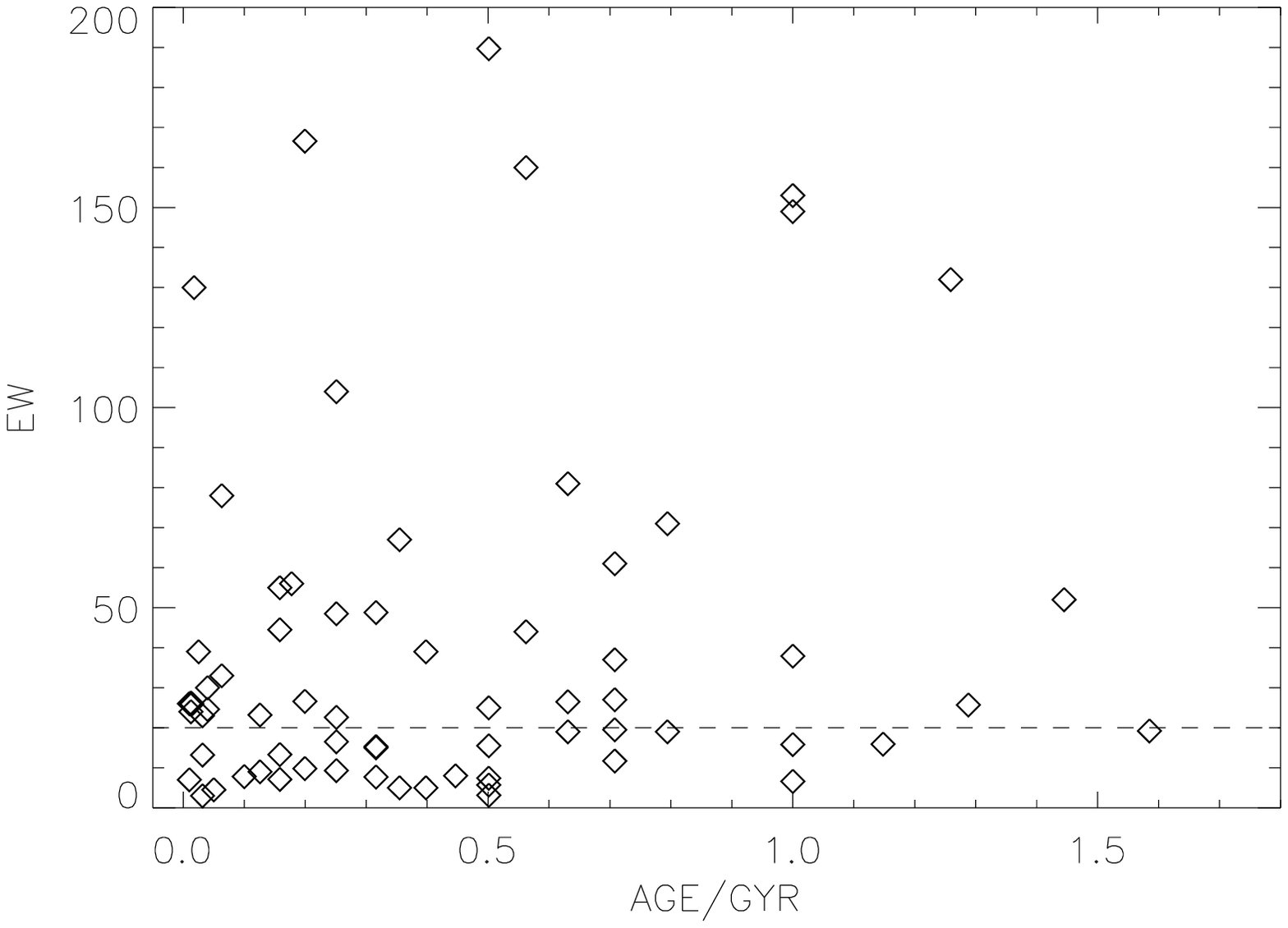}
\caption{Upper panel: the distribution of stellar ages for the LBGs 
in our sample. The full line refers to the best fit values, while the dashed line is the sum of the probability distribution function of all galaxies.
Lower panel: the dependence of stellar ages (best fit values) 
from the \lya\ EW. The dashed line is the 20 \AA value that is normally used for the narrow band selection of \lya\ emitters.
}
\label{fig1}
\end{figure}
In Figure 3 (upper panel), 
we show the derived stellar ages for our galaxies: the black 
histogram represents the distribution 
of best fit values from the SED fit for the entire sample, while the  
hatched green region
refers to the  NB subsample. 
In the lower panel we show how the age correlates with the \lya\ EW.
The age distribution presents a  peak at small values, with ages of a few tens of Myrs. The distribution drops quite quickly: about half 
of  the entire sample has ages shorter than 300 Myrs. 
The histogram then  shows the presence of a group of relatively old galaxies, 
peaked around $T=700$ Myrs and extending up to $\sim$1.5 Gyrs.
From the lower panel  we see no dependence of age on the EW.
The median ages are 350 Myrs for the entire sample and 
300 Myrs for  the NB subsample, but the difference is not 
statistically significant.
\\
A  comparison to values found in the literature is not straightforward, 
also because in many studies the 4000 \AA / Balmer break in not well sampled
 and therefore the usual dust/age degeneracy cannot be resolved. 
Age estimates for LBGs at z$\sim 5$ and $\sim 6$ are of the same order 
as those found in this study (e.g. Verma et al. 2007, Yan et al. 2007).
For LAEs, most authors estimate younger ages, of at most few million year 
(Pirzkal et al. 2007, Finkelstein et al. 2007, Malhotra \& Rhoads 2002) giving
support to the idea that LAEs are primitive objects.
However in some cases much  older stellar populations give equally good fit to the SEDs (Lai et al. 2007, Nillson et al. 2007, Gawiser et al. 2006). 
Recently Finkelstein et al. 
(2008b) reported that 2 out of their 15 LAEs have ages of several hundreds Myrs. 
\\
Since usually  the \lya\ emission, and in particular the  bright emission, 
with $EW> 100\AA$ is related to very young stellar ages 
(e.g. Charlot \& Fall, see also discussion in section 4), we have further 
checked the reliability of our results, in particular the validity of 
the fits giving old
ages. We have therefore derived the distribution of stellar ages, 
using for each object
 the probability distribution 
function (PDF) instead of the best fit value.
In practise for each object  we scan the parameter space  
 and determine the probability of each  particular 
model from the $\chi^2$ of the fit, 
as  $e^{-\chi^2}$.
The redshift of each galaxy is of course fixed to the spectroscopic one.
For each galaxy, we then compute the PDF
for all the physical parameters by scanning the $\chi^2$ levels
obtained during the fitting process. The 
probabilities of all models are normalized,
so that the sum is unity.
Then the range of values of the physical parameter analyzed 
(age, in our case) are scanned from a minimum value till
a maximum and we sum the probabilities  for all objects that fall within each  
particular interval, ending up with the PDF of the 
physical parameter for the entire sample.
\\
In this way we have derived the overall 
age distribution  using the sum of the  PDFs for each galaxy, 
for our full sample of 68 objects: in Figure 3 this is shown as a red 
dashed line. 
We see that the probability 
 distribution is in general agreement with the 
best fit one, but  is 
slightly shifted towards smaller ages.
We further checked the origin of this difference and we found that 
while the PDFs for all young galaxies 
( best fit age $T_{BF} < 200$ Myrs) 
are peaked around the best fit values with little dispersion, 
for some of the old galaxies (with $T_{BF} > 500 Myr$) 
the PDF  extendes also to much lower ages:
In practise the SEDs of  these galaxies could be fit almost equally well 
by models with much younger stellar populations and the age is not well 
constrained.
\\
To create a robust sample of ``old'' galaxies, we therefore selected 
 galaxies  with best fit ages $T_{BF} \ge 500 Myrs$, and minimum age for 
a reasonable fit of $T_{min} \ge  350 Myrs$.
In total there are 13 such galaxies in the entire sample.
 Any reasonable model that fits their SEDs must have a stellar population with age of 
at least few hundreds Myrs. Therefore
these galaxies are most certainly {\it not primeval galaxies} 
even if they show Ly$\alpha$ in emission.
Interestingly these old  galaxies have  values of Ly$\alpha$ EW  
that span the entire range from 3 \AA\ to 150 \AA\ as can be seen in Figure 3.
 Of the 13 old galaxies, 7 have $EW>20\AA$ and therefore they are part of the NB subsample.
In Figure 4 we show the SED of one of these bright emission 
line galaxies with old age, a galaxy at redshift 4.1: 
the best fit model is shown with a black line 
(best fit age 1.1 Gyr). The relative best fit models with younger ages 
(with age set equal to 10, 100, 200 and 600 Myrs respectively) are also 
shown with different colors. They 
clearly give a much poorer representation of the observed SED, 
expecially in the mid-IR range. 
Another object, from our sample of old galaxies with 
brigh Ly$\alpha$ emission, was already recognized 
by Wiklind et al. (2007) as an evolved and massive high redshift  galaxy 
(object number  5197 of their Table 4), 
since it has a prominent Balmer break. 
Although they did not have the spectroscopic redshift (z=5.56), 
their best fit model with a zphot of 5.2  gives results that 
 are entirely consistent with our fit: 
in particular their  best fit age and mass agree very well with ours.
The nature of these
old Ly$\alpha$ emitters will be further discussed in Section 5.
\begin{figure}
\includegraphics[width=9cm]{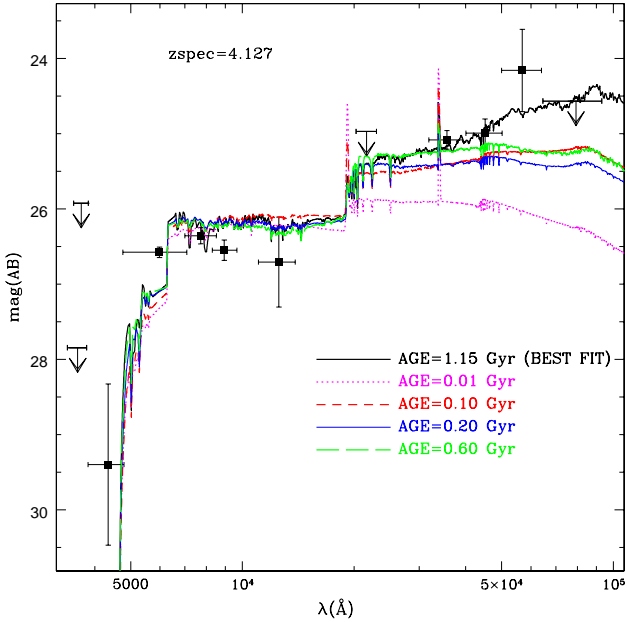}
\caption{The SED of galaxy with ID8073 at z=4.127: the black line is the best fit with age =1.1 Gyrs. The other lines with
different colors and types represent the bestfit models with varing ages. 
}
\label{fig4}
\end{figure}
\subsection{Lack of massive  LBGs with high EW Ly$\alpha$ emission}

In the upper panel of Figure 5 
we show the distribution of the  total stellar masses: the black line represents the best fit values while the red dashed line is the sum of the PDFs for each individual  galaxy, computed as in the previous section. 
The two agree very well with each other.
\\ 
The total masses range from  $10^9 M_\odot$ to as large as $10^{11} M_\odot$  with a  median value of 6$\times 10^9 M_\odot$. 
 The median  mass for the NB subsample is 
 5 $\times 10^9 M_\odot$. Similar values for z$\sim 5$ V-dropouts were obtained by Verma et al. (2007), and for $z\sim 6$ 
i-dropouts in GOODS survey by Yan et al. (2007); these last authors used 
also deep SPITZER-IRAC data to constrain the masses.
LAEs have in general much smaller masses: 
Finkelstein et al. (2007) find masses between $2\times 10^7 - 2 \times 10^{9} M_\odot$ for a sample of 98 LALA galaxies at redshift $\sim$4.5.  Gawiser et al. (2006) found an average mass per object of $\sim 5 \times 10^{8} M_\odot$ for a numerous sample of LAEs although at smaller redshift z$\sim$ 3.1 (see also Gawiser et al. 2007).
 Pirzkal et al.(2007) find masses between  $6\times 10^6 - 3 \times 10^{9} M_\odot$ for z$\sim 5$ LAEs from PEARS. Finally the most recent estimate comes from Finkelstein et al. (2008b) which show a larger mass range, from 
$10^8  M_\odot$ to as high as $ 6 \times 10^9 M_\odot$.
\\ 
More similar masses come from 
Lai et al. (2007) who find  $M=10^9 - 10^{10} M_\odot$ for 
three Spitzer detected  LAEs at z=5.7 in the GOODS north field.
The same authors in another study at redshift 3.1 of LAEs in the ECDF south, 
find that the median mass of LAEs is low, $ 3 \times 10^8 M_\odot$,
but the IRAC-detected LAEs (which make up 30\% of the sample)
 have masses of the order of $\sim 10^{10} M_\odot$.
\\
Clearly most (but not all) of the NB emitters have somewhat 
fainter continuum  than our LBGs: we are studying objects that are, on average,
 intrinsically brighter and thus more massive.
However the difference in mass is larger than expected: 
our sample comprises also 
galaxies with similarly faint broad band magnitudes,
thanks to the very deep  GOODS observations. 
For example the NB emitters of  Finkelstein et al. (2007) at z$\sim 4.5$
have typical i-band brighter  
than 26(AB),  while several of our V-dropouts, at a similar redshift, 
are even fainter than  this limit, in i-band. 
We also point out that most of the previously reported mass values, 
either lacked good IR data (which are of extreme importance for a reliable mass estimate) or were performed on stacked photometry.
\\
The difference  could be due, in part,  to the observed 
lack of massive galaxies with bright EW which is clearly seen in   
the lower  panel of Figure 5. Here we show how the stellar mass and 
\lya\ EWs  are related. The dashed line indicates the median mass of the 
entire sample, derived above.
Although we see no definite correlation between total 
stellar mass and EW,  we notice that 
while lower mass objects  span the whole range of EW from 
0 to 200 \AA , more massive objects 
have in general smaller EWs.
In particular if we take those  galaxies with $EW> 80 \AA$ (9 in our samples)
they all have masses that are equal or below the median mass.
\\
This effect is similar to  the deficiency of bright galaxies with large EWs 
that was initially  noted by Ajiki et al. (2004)  and recently confirmed by 
 Ando et al. (2007)  in a smaller sample of LBGs at redshift  
z$\sim 5-6$: they found that luminous LBGs 
(with absolute magnitudes $M_{1400}=-21.5$ ) generally show weak emission 
lines, 
while fainter LBGs show a wide range of \lya\ EW.
We also checked our EW as a function of absolute luminosity at 1400 \AA\, and confirmed the 
results of Ando et al. in terms of continuum luminosity, with a much larger sample of galaxies and  spanning a  slightly wider redshift range. This same result is confirmed by Vanzella et al. (in preparation) for all LBGs in the GOODS sample.
\\
This  effect cannot be due to selection biases, since bright/massive  
objects with large EW would not be missed in a spectroscopic survey
(on the other hand observations could  miss faint-small
 galaxies with small EW). 
Moreover it cannot be due to statistics given the size of the sample:
the total absence of such strong emitters in the more massive half 
 of the sample would be an implausibly large fluctuation.
We will discuss the possible origin of this effect  in Section 5.
Here we  point out  that this observational trend could partially  
explain the difference in mass between our sample and the sample of LAEs.
As already noted previously, NB surveys tend to select galaxies 
with extremely high EW. Although nominally the limit is 
20\AA, most objects have much larger EW, of the order of 100\AA or higher
(see for example Figure 1 of Finkelstein et al. 2007).
Thus they would naturally tend to select less massive galaxies than 
the LBG selection criteria, even in samples with similar broad 
band detection limits (e.g. see Figure 6 of Dijkstra \& Wyithe (2007)).
\begin{figure}
\includegraphics[width=9cm]{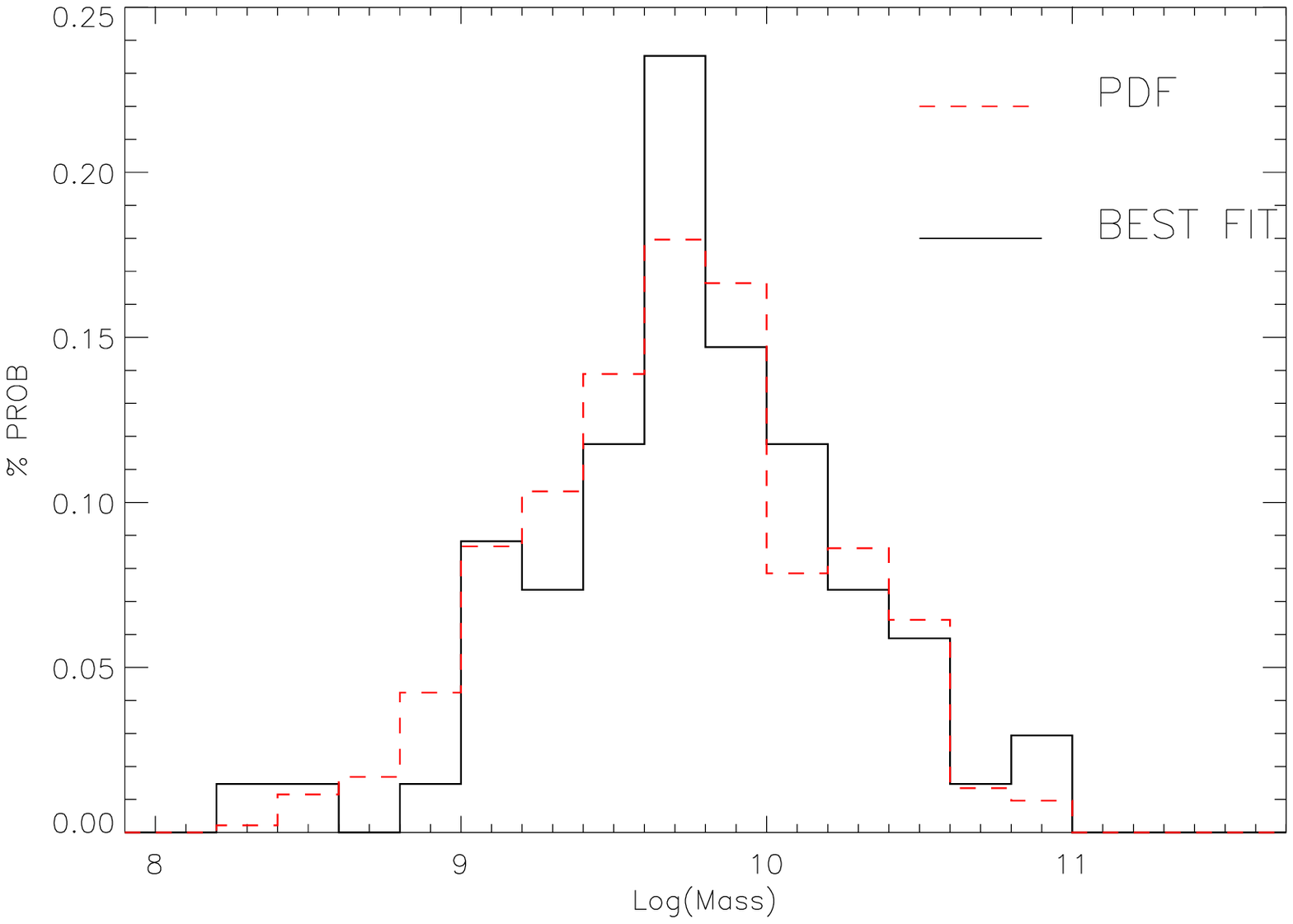}
\includegraphics[width=9cm]{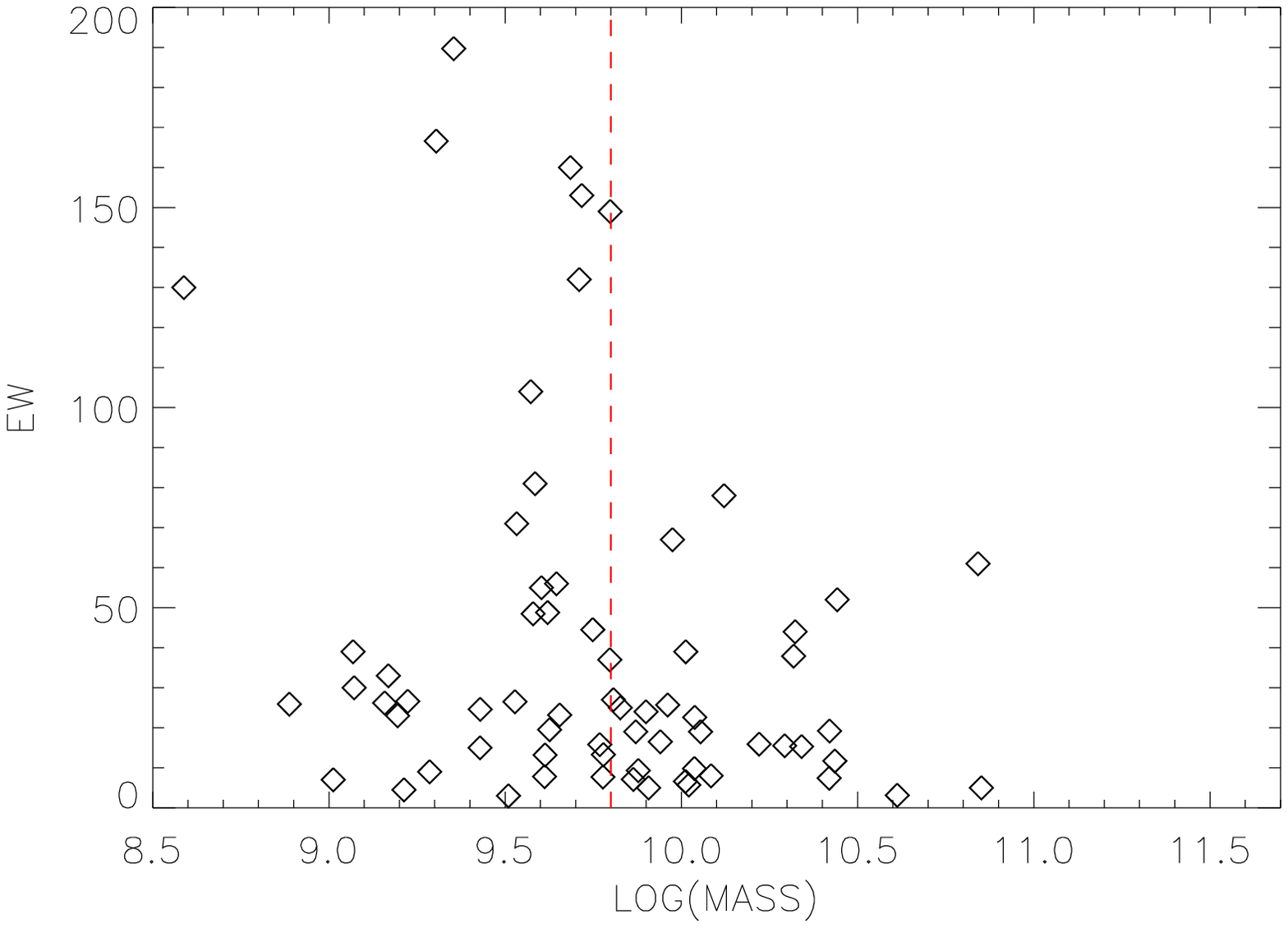}
\caption{Upper panel: the distribution of stellar masses for the LBGs 
in our sample (best fit values)
Lower panel: the dependence of stellar masses  (best fit values) 
on  \lya\ EW. The dashed line indicates the median mass.
}
\label{fig1}
\end{figure}
\subsection{Star formation rates}
There are different ways of estimating the SFR in line emitting galaxies.
\\
$\bullet$ The first and most obvious is from the Ly$\alpha$   emission, using 
the  calibration from Kennicutt (1998)  for the H$\alpha$ emission line  
$SFR=7.9 \times 10^{-42} L_{H\alpha}  M_\odot yr^{-1} $, 
and assuming case B recombination 
which gives $L_{Ly\alpha}=8.7 \times L_{H\alpha}$ 
where all luminosities are expressed in units of ergs/s. 
This SF estimator is sensitive to instantaneous 
star formation, since the flux depends mostly on very massive 
stars ($M > 20 M_\odot$).
\\
$\bullet$  The SFR can also be estimated 
from the  UV continuum using the  Kennicut conversion: 
$SFR_{UV}= 1.4 \times 10^{-28}  L_\nu  M_\odot yr^{-1} $, where  $L_\nu$
 is the luminosity at rest-frame 1400 \AA\ 
in units of ergs per second per hertz. 
This relation  assumes a $10^8$ years timescales for a galaxy to reach 
the full UV luminosity, so for the youngest objects the conversion 
could underestimate the SFR.
In any case the line-derived SFR is a more instantaneous measure 
than the UV-derived one.
Both the Ly$\alpha$ and  UV continuum photons are highly sensitive 
to the presence of dust although in a different way (see next Section).
Moreover both SFR estimators are highly dependent on the IMF.
Note that our UV continuum luminosity at 1400\AA\ 
comes out of the SED fit so it
depends also on the model.
\\
$\bullet$ Finally we have the SFR  value derived from the fit. This 
is not independent of the UV derivation, but it   
depends more heavily on the model assumed for the star formation history, 
in our case the exponentially declining star formation rate 
with e-folding time $\tau$.

The  values derived 
are between few and few tens of $M_\odot yr^{-1}$ for $SFR_{Ly\alpha}$ and 
$SFR_{UV}$, in broad agreement with the range found e.g. by Tapken  et al.
 2007 for LBGs with Ly$\alpha$ in emission and  by Ajiki et al. (2003) for Ly$\alpha$ emitters at z=5.7.
\\
Like these authors we then examined how $SFR_{UV}$ and  $SFR_{Ly\alpha}$
compare to each other. 
In principle the ratio  $SFR_{Ly\alpha}/SFR_{UV}$ is directly related to the EW value. However for the present work we point out that 
EW and Ly$\alpha$  total flux are measured directly from 
the spectra, with  the reference continuum taken  at $\sim 1200 \AA$; 
the UV  continuum that goes into the  $SFR_{UV}$ determination  is
 derived from a fit to  the broad band photometry and at 1400\AA\ restframe.
Therefore the line EW and flux come from a region  of width $1''$ 
centered on the galaxy, corresponding to the slit 
width used  in the spectroscopic observations.
The UV continuum (and the SED fit) refers to  aperture photometry flux,
and can be assumed to be the ``total flux''. 
This should not be a big problem since  the Ly$\alpha$ emitters
are in general very compact galaxies, as shown by Lai et al. (2007).
Moreover LBGs with Ly$\alpha$ in emission are in general smaller 
than those with the line in absorption, as recently shown
by Vanzella et al. (2008 in preparation). 
Their typical half light radius is $\sim$ 5 ACS pixels i.e just 0.25$''$.
Therefore most of the flux should come from the 
inner region, well within the slit-width used for spectroscopy.
Clearly this refers to the continuum morphology 
but a comparison of NB  and continuum images for 
Ly$\alpha$ emitters around  several high redshift radio galaxies 
shows that the two are  very well correlated  (Venemans et al. 2007).
\\
We therefore estimated the  correction factor:
we assumed that the line emission is distributed as the continuum,
smoothed the high resolution continuum image (closest 
to the Ly$\alpha$ $\lambda$), to the ground based spectroscopic 
resolution and then estimated the flux inside 
 the slit aperture. We repeated this for all galaxies and
found that the fraction of flux that falls inside the 
slit is $\sim$ 80\% of the total. 
\\
The results are presented  in Figure 6.
Note that all these values are uncorrected for dust extinction, 
so the true star formation rates are likely to be higher.  
\\
The values are comparable to what is found for LBGs at similar redshifts  
by e.g.  Tapken et al. (2007), Stanway et al. (2007) and Verma et al. (2007)
 and  for LAEs at z=3-6  (Ajiki et al. 2003, Venemans et al. 2005). 
\\
The median  ratio $SFR_{Ly\alpha}/SFR_{UV}$ is $\sim 0.7$, so in general the $
SFR_{UV}$ is larger than the  $SFR_{Ly\alpha}$ but the scatter is very large.
Similar results although with large variations were found both for LAEs 
and for LBGs with $Ly\alpha$ emission.
For continuum selected high redshift galaxies,
 Tapken et al. (2007) find a median  
ratio of  0.2 with values ranging from 0.1 to 20  while at higher redshift 
Dow-Hygelund et al. (2007) find values between 0.27 and 1.2 with a 
large scatter 
for i-dropouts. For \lya\ emitters 
Ajiki et al (2003) find a median   $SFR_{Ly\alpha}/SFR_{UV} \sim 0.5$; 
whereas Venemans et al (2006) 
find  a mean of 0.6-0.7 for galaxies  in a protocluster around a radio-galaxy at redshift 3.1
 \\
The lower average value of  $SFR_{Ly\alpha}$  are therefore  ubiquitous and can  in general be  attributed to the effect of 
dust extinction and scattering by the intergalactic medium. However, 
some of our galaxies have  $SFR_{Ly\alpha} / SFR_{UV}$  larger than  1. These 
galaxies could be  in the  very early phases  of star formation activity, 
in which  $SFR_{UV}$  values are underestimated (Schaerer 2000; 
see also Nagao et al. 2004, 2005). 
Indeed by checking the best fit ages estimated from the multiwavelength SED, 
we see that most of these galaxies having  $SFR_{Ly\alpha} / SFR_{UV} >$ 1 are extremely 
young, with ages of a few tens of Myrs.
As illustrated in Figure 6 (bottom panel), objects with small $SFR_{UV}$ have 
higher $SFR_{Ly\alpha} / SFR_{UV}$ ratios, i.e. these very young galaxies tend to have also relatively modest SFR values.
\\
The wide variety of  $SFR_{Ly\alpha} / SFR_{UV}$  could also be the effect of
the different way in which dust suppress Ly$\alpha$ photons and continuum photons, which will be discussed in the next subsection.
\begin{figure}
\includegraphics[width=9.5cm]{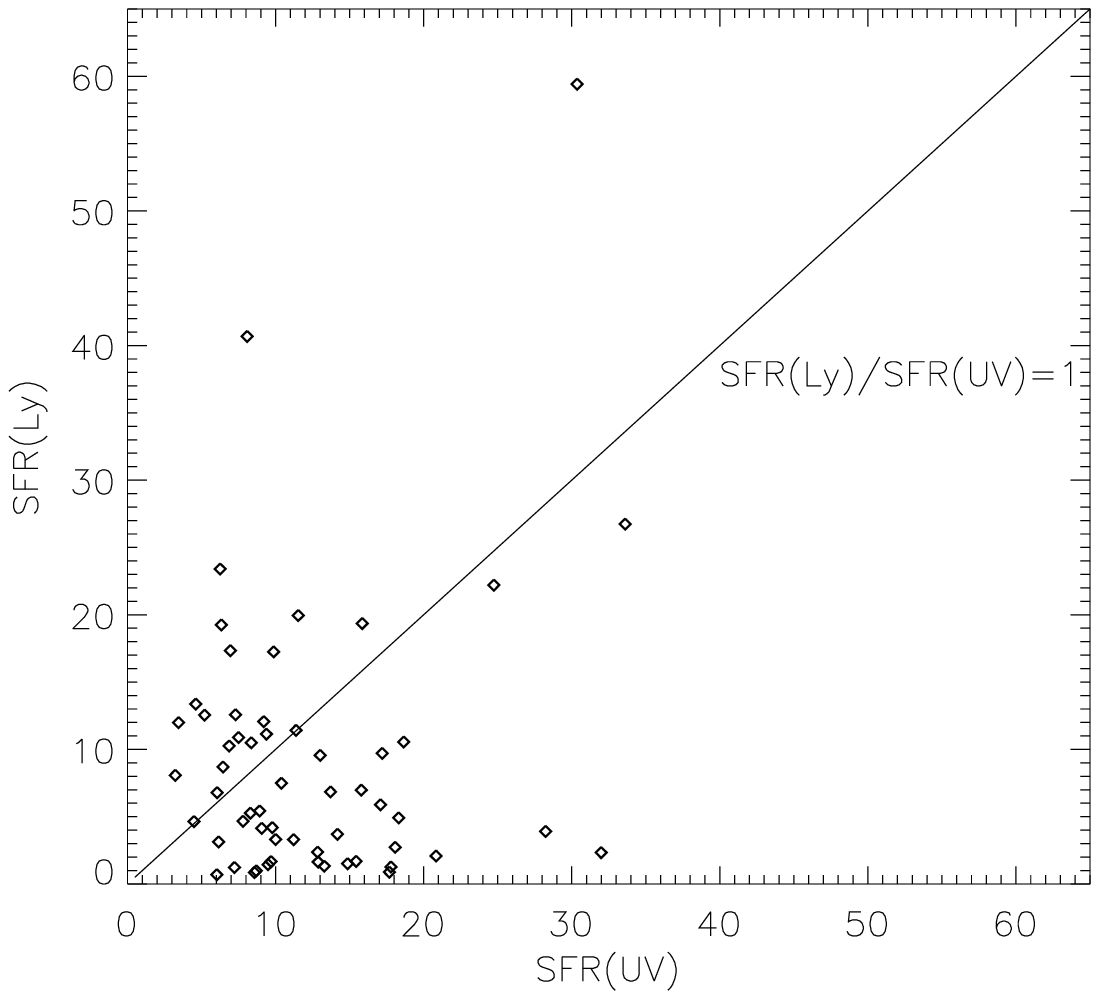}
\includegraphics[width=9.5cm]{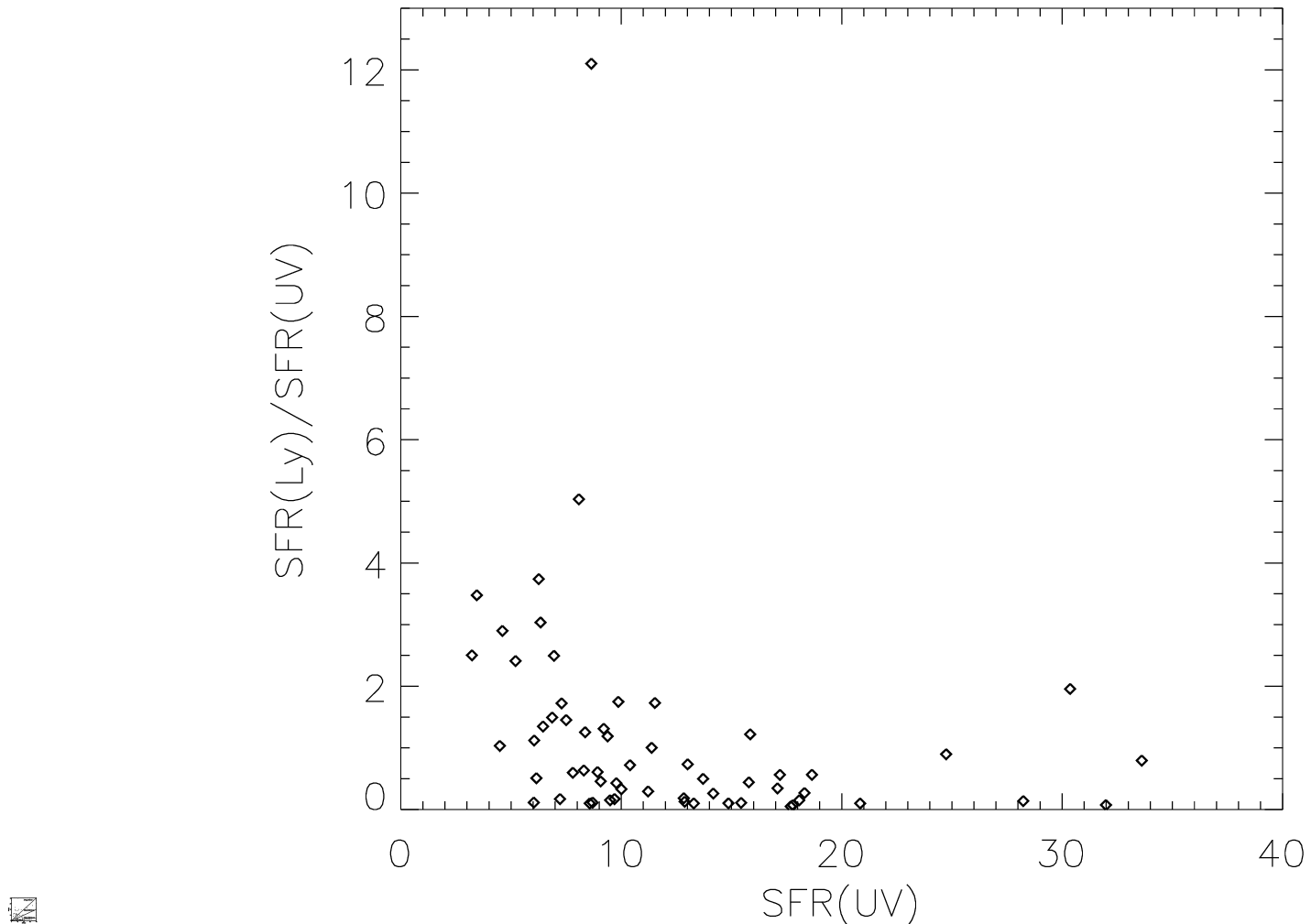}
\caption{ Upper panel: the star formation rate derived from the UV continuum at 1400 \AA  versus the SFR derived from the \lya\ line emission (uncorrected for dust extinction, see text for details). Both values are in $M_\odot yr^{-1}$.
Lower panel: the ratio SFR(\lya)/SFR(UV) versus the total SFR(UV).
}
\label{sfr}
\end{figure}

\subsection{Dust}
While in general Ly$\alpha$ emitters are regarded as a 
dust free population (e.g. Lai et al. 2008, Gawiser et al. 2006),
we find that the presence of a modest 
but non-zero amount of dust is required by the SED fit of many galaxies.
The fitted E(B-V) parameter is not zero in about 2/3 of the galaxies, 
with individual galaxies showing values as high as $A_v \sim 1$.
However the  mean extinction of the whole sample is very low,
corresponding to $A_v \sim 0.25$ (roughly $A_{1200}\sim 1$). 
This fits in the trend recently noticed by Finkelstein et al. (2008b),
that studies which analyze objects separately seem to detect dust extinction
(at least in some galaxies), while those that stack fluxes do not.
\\
In our previous  paper (P07) we found that LBGs with \lya\ in absorption 
 are actually dustier than the LBGs with \lya\ in emission: they had redder   slope $\beta$, as  determined directly from the 
observed colors i-z  and  consequently had,
on average, a higher   E(B-V) parameter, as derived from the spectral fitting.
However, the differences we found were  not very large, possibly  due the small size of our sample.
With our enlarged  sample  
we checked if there is any dependence on the E(B-V) parameter on the  \lya\ 
equivalent width, by dividing our sample into 4 sub-groups
according to the value of E(B-V), respectively E(B-V)=0,0.03,0.06-0.1 0.15-0.4. 
For each group we determined the mean EW, and we plot them in Figure 7 
(error bars for EW indicate the standard error of the mean, while the bars in E(B-V) 
correspond to the range of values in each subsample).
As we can see there is a net trend of \lya\ EW with dust extinction, with 
unextincted galaxies 
showing on average higher \lya\ EW, similar to what found by Shapley et al. (2003). The number of 
objects is approximately equal in each subgroups, therefore the larger error bars for the objects 
with less exctinction 
reflect a larger dispersion of the values around the mean.
\\
Previously Shapley et al.(2003) also  found similar trend between E(B-V) and EW, for 
LBGs at z$\sim 3$. From composite spectra they found
 considerable  difference between the $LBGs$ with and without \lya emission,  
with the former having a steeper slope. They also found a significant positive
dependence of slope  on the \lya\ equivalent width.
\begin{figure}
\includegraphics[width=9.5cm]{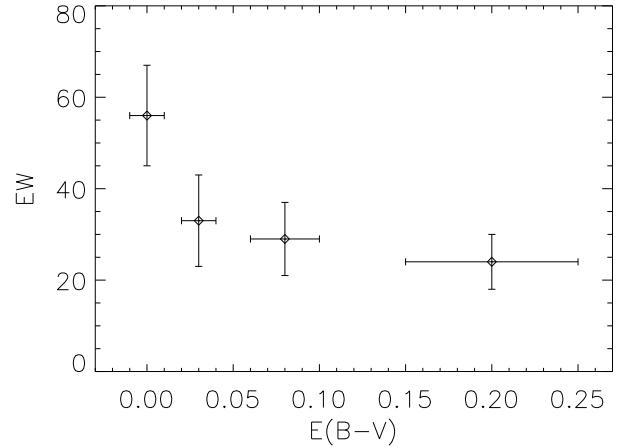}
\caption{The average EW of galaxies divided by values of E(B-V).Error bars for EW are  
the standard error of the mean, 
error bars in E(B-V) correspond to the range of values in each subsample
}
\label{sfr}
\end{figure}
Finally Vanzella et al. (2008 in preparation) analysed composite spectra of B, V and i-dropouts from GOODS: 
in particular for the z$\sim 4$ sample the composite spectrum of the absorbers has a   spectral slope considerably redder than the emitters, in accordance with our results.
\\
Finkelstein et al.(2007, 2008a) proposed an interesting scenario in which 
dust effects could enhance the Ly$\alpha$  EW  by allowing the \lya\ 
photons to escape, even if the continuum is extinguished.
In other words  dust can selectively suppress the continuum emission
 but not the \lya: this is possible in a clumpy  
medium where dust is primarily in cold neutral clouds, whereas the inter-cloud medium is hot and mainly ionized (e.g. Hansen \& Oh 2006). 
Since we find that in general EW and E(B-V) are well correlated, and the larger the E(B-V) inferred from the continuum, the smaller the Ly$\alpha$  EW,   
we can conclude that 
on average dust suppresses in a similar way 
both the continuum and the \lya\ photons.  Therefore
the clumpy scenario is not needed. 
However we cannot exclude that it can work for some individual galaxies.
We have not attempted a full SED modeling of our galaxies
 to include  the clumpiness {\it q} parameter 
{\it a  la} Finkelstein et al., but we can search our sample for galaxies 
with non negligible dust content,
and at the same time a relatively bright Ly$\alpha$ emission.
\\
We find one galaxy at redshift 5.5 
which shows a bright \lya\ line, with  
$EW \sim 80 \AA$ (rest-frame)
and a relatively large value of E(B-V)=0.4. 
We checked the SED and it is indeed a quite red and massive galaxy 
($M \sim 10^{10} M_{\odot}$). Its slope in the  near-IR rest-frame  
is well constrained by detections in the four IRAC bands. 
The observed $SFR_{UV}$ is also about 3.5  higher than the $SFR_{Ly\alpha}$.
It is therefore possible 
 that in this galaxy a clumpy dusty ISM  could  enhance the EW.
Actually Finkelstein et al.  argue that this mechanism  could be at work 
also in galaxies with smaller amounts of dust:  they find few  
objects  that require values of $A_{1200} =1$ (corresponding to $A_v =0.25$
 approximately) but where the dust enhancement is necessary.
 A full modeling of our galaxies
within this scenario is beyond the scope of this paper and
 is deferred to future work.

\subsection{New Charlot \& Bruzual models}
All previous results have been obtained with the  
Bruzal \& Charlot models (BC03), since these are the most used 
models in  the comunity, and we wanted to make a direct comparison to previously published  work.
In this subsection we briefly summarize the results obtained with the more 
recent CB07 models.
The obtained masses are on average  20\%  lower then those obtained with 
the BC03 models
and the ages are younger also by an average  20\%, while the continuum derived star formation rates are more similar.
Moreover the differences are  systematic and do not present a large scatter; 
this implies that qualitatively the main results (i.e. the presence of old and massive Ly$\alpha$ 
emitters ) do not change even using the new libraries, but are only mildly rescaled.
As an  example if we consider the ``solid'' old galaxies  with the same requirement as in section 3 
(i.e. best fit age larger than  500 Myrs and minimum age for a good fit $\geq$ 300 Myrs) we find 
11 (instead of 13) galaxies with only two objects dropping out of the group.
All correlations between properties that  we have presented, namely the age vs EW,  total 
mass vs EW and E(B-V) vs median EW,
do not change with the new models.
\\
Note that the differences in mass and age that we find are somewhat smaller 
than those reported in the literature, e.g. 
a 50\% mass difference reported by Bruzual (2007).
This is not surprising since our sample is at high redshift.
In the  new models (as well as those of Maraston 2005)
the most notable change is the inclusion of the TP-AGB 
phase of stellar evolution: its  contribution,  in the 0.2-2 Gyrs age range,  
is at a maximum in the near-IR (K band) rest-frame  which we  do not sample
even for the lower redshift galaxies in our sample.

\section{Discussion and conclusions}
We have analyzed a sample of 68 Ly$\alpha$ emitting LBGs 
analyzing their physical and spectral properties.
Here we summarize the main results and discuss them:

 Although most galaxies are fit by  young stellar populations, 
a small but non negligible fraction
has SEDs that cannot be well represented by young models and 
require considerably older stellar component  up to $\sim$ 1 Gyr.
Age and EW do not show a strong correlation. Some of the 
robust  ``old'' galaxies have EW as high as 100 \AA and therefore 
in principle they should be present also in narrow-band selected 
samples of LAEs.
\\
The  presence of these old galaxies with strong \lya\ emission
 is also important for modeling the entire LAE population. 
For example most authors that have modeled the Ly$\alpha$ population  
to reproduce the  LF and clustering properties of LAEs and of LBGs 
(e.g. Mao et al. 2007, Mori \& Umemura 2006, Thommes \& Meisenheimer 2005) 
assume in general much shorter timescale for Ly$\alpha$ emission
of the order of  $\sim 1-200$ Myrs at most.
\\
Recently Finkelstein et al. (2008b) found 2 out of 15 LAEs to be consistent with evolved galaxies with ages around 0.5 Gyrs, with the rest confined to ages below few tens of Myrs. They suggested a possible bi-modality in the age distribution of LAEs. 
They also suggested that the clumpy dusty ISM scenario (already discussed above) could cause old galaxies
 to still have Ly$\alpha$ in emission. 
\\
Similarly   Thommes \& Meisenheimer (2005) presented model calculations on the 
\lya\  emitting primeval galaxies.
They suggested  the possibility of a double phase activity 
for the \lya\ emission, i.e. that the  \lya\ had a initial bright phase 
with a  short timescales,  
due to primeval gas in almost  dust-free galaxies, and a secondary phase 
at much later time. 
\\
As a possible test to this model, they predict that in the galaxies undergoing the second bright phase of emission, 
the dust has been swept away by gas outflows, 
so the \lya\ lines should be shifted  by the velocity of 
the outflowing wind relative to the metal absorption lines 
and should show a P-Cygni profile. 
Furthermore the relative contribution of the second generation of these bright \lya\ emitters should  increase 
with cosmic epoch, i.e while at redshift $\sim  5-6$ most of the \lya\ emitters should be primeval galaxies,  at redshift 3, i.e. a Gyr after, there should be also a lot in this second bright phase. 
\\
Both this effects could be tested. In a forthcoming paper we will present 
results on a sample of  LBGs with \lya\ emission at z$\sim$2.5-3 from the 
GOODS-south field,  to see  
if the age distribution of these galaxies is significantly different 
from the present sample.

 We then found a  lack of galaxies with large stellar masses and large EW, 
which  
cannot be due to a selection effect or a statistical fluctuation.
There could be several possible causes for this trend: a first 
possibility is that  the brightest/most massive  
galaxies might reside in more dusty environment compared to less 
massive  galaxies.
We do not find significant differences in the E(B-V) 
values between  massive  galaxies and the rest of the sample, 
but we cannot exclude this possibility.
 \\
The amount of HI gas in and surrounding the galaxies could also affect the \lya\ EWs:
the most massive galaxies probably reside in more massive DM halos, and they could be  
surrounded by a larger amount of HI gas that selectively
 extinguish the \lya\ emission, resulting in smaller EWs. This also fits in the 
biased galaxy formation scenario. 
 Finally as argued by Shapley et al. (2003), LBGs with smaller EW 
have in general also stronger 
LIS absorption and large velocity offset of the \lya\ emission: 
it could therefore be that these galaxies (which are most massive)
 contain more  outflowing neutral gas with a large velocity dispersion  
that would depress partially the \lya\ emission  resulting in a smaller EW.
This would result in more asymmetric line profiles, 
which could be observed  in higher resolution 
spectra (such as those by Tapken et al. 2007).

The presence of dust, although in small amounts, is required by the 
SED fit of many galaxies. 
Therefore Ly$\alpha$ emitters are not completely dust free galaxies.
The amount of dust and the EW are well correlated, at least on average.
This was already observed at redshift$\sim 3$  by Shapley et al. (2003) and at redshift $\sim 4$ in our previous work (P07).
 In a recent paper Schaerer \& Verhamme (2008) present the results of the application 
of their radiative transfer model (Verhamme et al. 2006)  on the well studied LBG 1512-cB58 at z$\sim$3.
From this analysis  they derive  the interesting 
implications that even to model the spectra of LBGs where Ly$\alpha$ is  present in 
absorption,  an intrinsic relatively high EW ($> 60-80 \AA$) Ly$\alpha$ line is required.
They propose that the vast majority of LBGs have intrisecally high EW (60 or larger) and that 
the main physical parameter responsible for the observed variety of line profiles 
and strengths in LBGs is the HI column density $N_H$, and the accompanying variation 
of the dust content.
This model explains naturally the trend of EW with E(B-V) parameter found in Figure 7.
It also explains the difference in the UV slope  parameter  between 
the LBGs with and without \lya\ emission, that we found in P07 for z$\sim 4$ galaxies 
and the analogous results by Shapley et al. (2004) at z$\sim3$.
Given the observed mass metallicity relation, it is natural to speculate 
that the most massive galaxies are in general also dustier: this could 
easily explain the lack of massive galaxies with very large \lya\ EW that we show in Figure 1,
but as already noted above we do not recover any net correlation between mass and E(B-V),
possibly due to the large scatter.

Finally we found that the $SFR_{UV}$ and  $SFR_{Ly\alpha}$  are 
similar, with the 
 $SFR_{Ly\alpha}$ on average somewhat lower than the other values, 
in agreement with previous results.
There is a large scatter in the values of  $SFR_{Ly\alpha}/FR_{Uv}$ for 
individual galaxies which  could be due 
to the different effect of dust on the continuum and \lya\ photons or to other effects, such as the variations in the gas metallicity or fluctuations in the opacity of the IGM.
\\
\\
In conclusion, there seems to be  a continuity of 
properties between LBGs with faint \lya\ emission, and those with brighter 
\lya, such that they would be selected also by NB searches. 
Not all \lya\ emitting galaxies
 are small, young dust free galaxies. 
A non negligible fraction shows older stellar 
populations, with ages up to $\sim$1 Gyr and masses in 
excess of $10^{10} M_\odot$. 
The lack of massive objects in NB selected samples can be partly explained 
by the observational trend presented in  Figure 5.
On the other hand, relatively 
old galaxies should be present also in NB selected samples.
This difference could be in part ascribed to two effects: first, most of the 
NB studies do not have a wavelength range as wide as GOODS 
to effectively determine the physical properties, 
and/or lack data in the fundamental region 
around 4000\AA\ break which  are essential to reduce model
 degeneracies (although they cannot be solved completeley).
Second, and most important, we find that there is a large 
diversity of properties amongst \lya\ emitting galaxies. Therefore, those 
studies that rely on stacked photometry to derive the average physical 
properties  might miss to represent the galaxies with the most extreme
 characteristics: e.g. the presence of a 20\% old galaxies could be missed if 
``diluted'' in a more numerous, much younger population.

The present  study shows that the simple picture LAEs equal 
young galaxies,  LBGs equal older galaxies cannot explain 
all the observed properties and trends. 
A more complex scenario is probably needed, including variables such as 
the dust content, the ISM clumpiness, the amount and kinematics of neutral 
gas, and perhaps  the viewing angle of galaxies 
as recently suggested by Laursen et al.(2008).

\end{document}